\documentclass[11pt, a4paper]{article}

\usepackage{amsmath}
\usepackage{amsfonts}
\usepackage{amssymb}
\usepackage{graphicx}
\usepackage{mathrsfs}
\usepackage{epsfig}
\usepackage{latexsym}
\usepackage{graphicx}
\usepackage{color}
\usepackage{amsmath,amssymb}
\usepackage{hyperref}

\numberwithin{equation}{section}

\hypersetup{colorlinks, citecolor=bluscuro, linkcolor=black, urlcolor=bluscuro}
\definecolor{rossos}{rgb}{0.8,0.2,0.3}
\definecolor{bluscuro}{rgb}{0.15, 0.2, .85}
\definecolor{bluchiaro}{cmyk}{1,.3,0.,0.1}

\makeatother   
\newcommand{\GeV}{{\rm \,GeV}}
\newcommand{\TeV}{{\rm \,TeV}}

\def\de{\textrm{d}}

 \def\be   {\begin{equation}}   \def\ee   {\end{equation}}
 \def\ba   {\begin{array}}      \def\ea   {\end{array}}
 \def\bea  {\begin{eqnarray}}   \def\eea  {\end{eqnarray}}
 \def\bean {\begin{eqnarray*}}  \def\eean {\end{eqnarray*}}

\begin{document}

\begin{flushright} 
CERN-PH-TH/2013-054\\
SISSA  08/2013/FISI
\end{flushright}

\vspace{0.5cm}
\begin{center}

{\huge \textbf {
Interpretation of AMS-02 Results: \\[0.3cm]
Correlations  among Dark Matter Signals
}}
\\ [1.5cm]
{\large{\textsc{
Andrea De Simone$^{\,a, b, c,}$\footnote{andrea.desimone@sissa.it}, 
 Antonio Riotto$^{d,}$\footnote{antonio.riotto@unige.ch}, 
 Wei Xue$^{\,c, b,}$\footnote{wei.xue@sissa.it}
 }}}
\\[1cm]

\large{\textit{
$^{a}$ CERN, Theory Division, CH-1211 Geneva 23, Switzerland
\\  \vspace{1.5mm}
$^{b}$ SISSA, via Bonomea 265, I-34136 Trieste, Italy
\\  \vspace{1.5mm}
$^{c}$ INFN, Sezione di Trieste, via Bonomea 265, I-34136 Trieste, Italy
\\  \vspace{1.5mm}
$^{d}$ D\'epartement de Physique Th\'eorique and Centre for Astroparticle Physics (CAP),\\
24 quai E. Ansermet, CH-1211 Geneva, Switzerland
}}
\end{center}

\vspace{0.5cm}

\begin{center}
\textbf{Abstract}
\begin{quote}
The \textsc{AMS-02}   collaboration 
 has recently released  data on the positron fraction $e^+/(e^-+e^+)$ up to energies of about 350 GeV.
If one insists on interpreting the  observed excess as a dark matter signal, then we 
find it is best described by a TeV-scale dark matter annihilating into $\tau^+\tau^-$, although this situation  is already severely constrained by gamma-ray measurements.
The annihilation into  $\mu^+\mu^-$ is allowed by gamma-rays more than $\tau^+\tau^-$,
but it gives a poorer fit to \textsc{AMS-02} data.
Moreover, since electroweak corrections induce correlations among the  fluxes of stable particles from dark matter annihilations, 
the recent \textsc{AMS-02} data imply a well-defined prediction for the correlated flux of antiprotons.
 Under the assumption that their future  measurements will not show any antiproton excess above the background,
 the dark matter interpretation of the positron rise will possibly be ruled out by only making use of data from a single experiment. 
This work is the first of a program where we emphasize the role of  correlations among dark matter signals. 
\end{quote}
\end{center}

\def\thefootnote{\arabic{footnote}}
\setcounter{footnote}{0}
\pagestyle{empty}

\newpage
\pagestyle{plain}
\setcounter{page}{1}

\section{Introduction}
Despite the overwhelming evidence that Dark Matter (DM) 
contributes to a significant fraction of the energy density of the  universe,  its nature and properties are still unknown. 
Many  experimental activities are   nowadays performed assuming the particle nature of DM.  In particular, 
in  the  indirect searches of  DM \cite{review} the goal is to  detect the products of 
DM annihilations or decays
around the Milky Way: Standard Model (SM)   particles are produced as primaries and they subsequently 
decay into SM stable states such as electrons and positrons, protons and antiprotons, deuterium and antideuterium, 
photons ($\gamma$-rays, X-rays, synchrotron radiation) and  neutrinos.
In praticular, the recent  positron data  released by \textsc{AMS-02}
\cite{ams02positron} may be interpreted as due to DM annihilations.

The necessary    inclusion of ElectroWeak (EW)  
corrections \cite{dm1,dm2,dm3,dm4} significantly alters the final  spectra when the mass of the DM particles  is larger than the EW scale (see also \cite{dm5,dm6,dm7}): soft EW gauge bosons are copiously radiated, thus 
opening new channels in the final states which otherwise would be forbidden if
such corrections were neglected. As a consequence, all stable particles are  present in the final spectrum, independently of the primary channel of DM  annihilation/decay.  
An  inevitable consequence of  this is
inducing correlations among the predicted  fluxes of SM stable particles targeted by the various experiments. For example,  within a given model of DM,  EW corrections to DM annihilations/decays lead to correlated fluxes of antiprotons ($\bar{p}$) and neutrinos whose detection is one of the goals of \textsc{AMS-02} and \textsc{IceCube}, respectively.

This work is the first of a program where we investigate  the role of 
correlations among DM signals. Based on the idea that EW interactions connect all particles of the SM and that  correlations can be established among different fluxes, our logic is very simple:
\begin{enumerate}
\item suppose a signal is detected in a particular channel and assume
it is of DM origin;
\item then the EW corrections imply that the fluxes of other cosmic-ray
species should be modified accordingly, in a very specific way depending
on the nature of the process which is assumed in part 1;
\item these correlations can then be tested against existing data, or
used as predictions for upcoming data,
to verify or reject the initial hypothesis of DM interpretation of the signal.
\end{enumerate}
This 
 approach is essential  to draw robust conclusions about DM origin of a signal, and it is
 particularly relevant now, due to the amount
of upcoming data from very different sides and channels (e.g. \textsc{AMS-02},
\textsc{Fermi}, \textsc{HESS-2}, \textsc{IceCube}, \textsc{Pamela}).

In this paper we apply the general logic described above to the specific case of the \textsc{AMS-02} experiment,
which has recently released positron data \cite{ams02positron} and will release data from anti-proton measurements
in the very near future. 
We first consider the recent \textsc{AMS-02} data on the positron fraction $e^+/(e^++e^-)$  (which extend to higher energies the observations by \textsc{Pamela} \cite{Adriani:2008zr} and 
\textsc{Fermi}
\cite{FermiLAT:2011ab}) and fit them as a signal of DM.
Then, 
the EW corrections provide a prediction for the  correlated flux  of $\bar{p}$ (which are 
already  constrained by current \textsc{Pamela} data \cite{Adriani:2010rc}). Once this prediction is made, we may   compare it with the projected  sensitivity of   \textsc{AMS-02} for  the $\bar{p}$ channel  under the assumption   that  the  measurements of \textsc{AMS-02} do not show  a significant $\bar{p}$ excess  above the background.  

 The correlation between $e^{+}$ and $\bar{p}$  is  particularly relevant as it  is
 a very clear example to illustrate the role played by EW corrections in constraining DM models even within a  single experiment, such as \textsc{AMS-02}, which is able to measure the fluxes of different particle species.
 The inclusion of EW corrections will allow us to answer a simple and interesting question: will it be possible
 to rule out conclusively  the DM interpretation of the positron rise
using only the data collected by the \textsc{AMS-02} experiment  (and thus disregarding other experiments)?     
Notice that the correlations between different final states can also help to break the degeneracy implicit in the interpretation of the $e^{+}$ signal from either DM or astrophysical sources, mainly pulsars. While it was recently concluded that future $e^{\pm}$ data will likely be insufficient to discriminate between the DM  and
the single pulsar interpretations of the cosmic-ray lepton
excess \cite{bertone}, this conclusion did not take into account EW corrections and the correlations induced by them. 
The correspondingly generated flux of $\bar{p}$ might be detected by \textsc{AMS-02} and not be ascribable to pulsars. We will come back to this point in a separate publication  \cite{inprep}.

The paper is organized as follows. 
In Section \ref{sec:background} we  briefly discuss the signals and the backgrounds
relevant for our analysis.
Our results  are presented and discussed in Section \ref{sec:results},
including a description of the salient features of our detector simulation.
The concluding remarks are collected in Section \ref{sec:conclusions}.

\section{Signals and backgrounds}
\label{sec:background}

As a signal, accounting for the positron excess observed in the data, 
we consider DM annihilations in the following leptonic channels
\be
\textrm{DM\,\,DM}\to  e^+ e^-, \, \mu^+\mu^-,\, \tau^+\tau^-,\,
\label{channels}
\ee
each with 100\% branching ratio. We will not consider the case of decaying DM.
The above annihilation channels are the best  
situations for interpreting the positron rise as a signal of DM.
Other non-leptonic annihilation channels, such as $b\bar b$ and $W^+W^-$,
easily produce hadrons in the final state, e.g. $\bar{p}$,
and are already excluded by the $\bar{p}$ \textsc{Pamela} data \cite{Adriani:2010rc},
using EW corrections \cite{dm1}.
Note however that other complimentary data put already strong constraints on these
``leptophilic'' scenarios,  e.g.  $\gamma$-rays  
\cite{Meade:2009iu, Cirelli:2009dv}; as mentioned above, we insist on exploring
the exclusion capabilities of 
data from a single experiment, with EW corrections taken into account.

\begin{figure}[t!]
\centering
\includegraphics[scale=0.4]{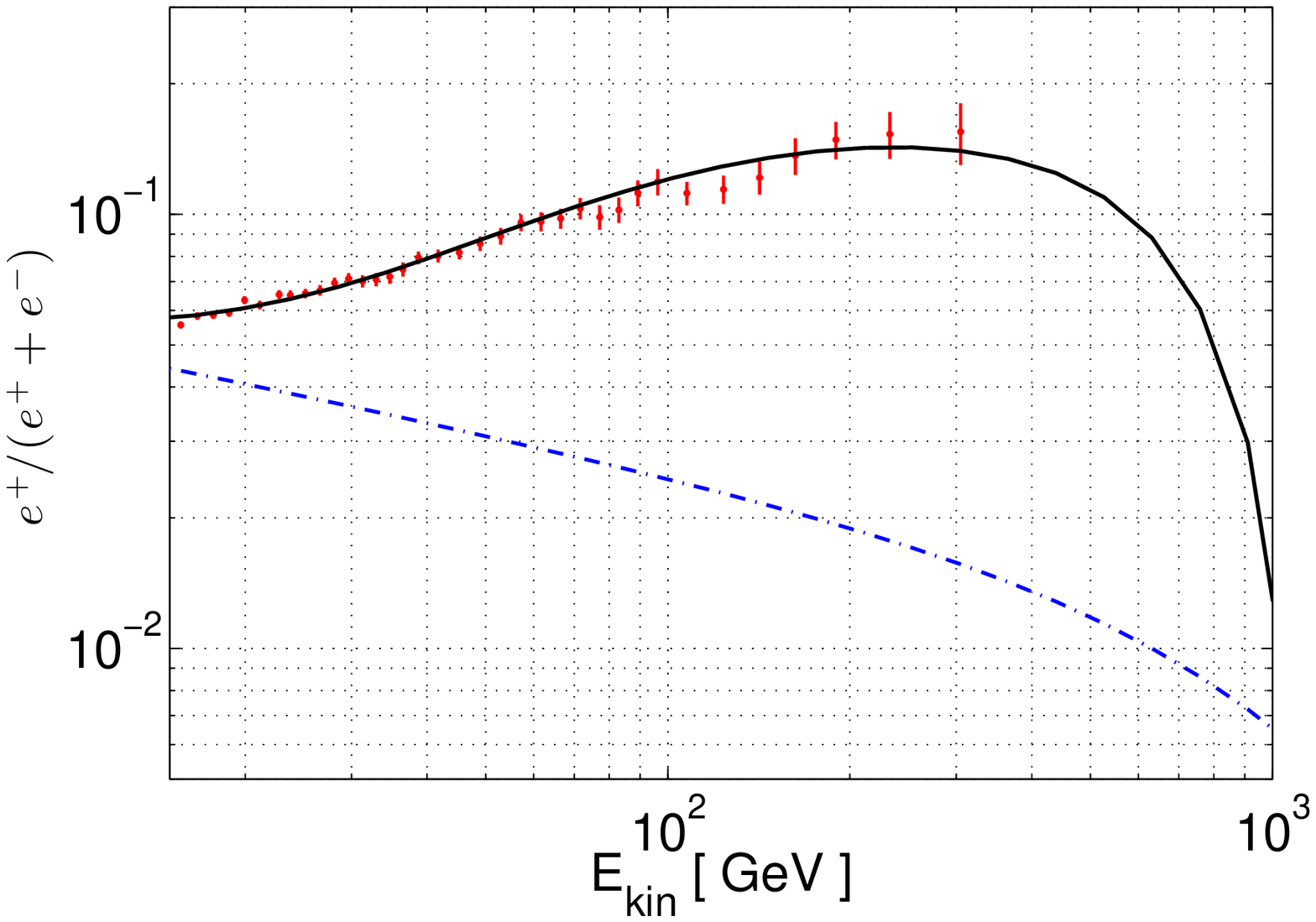}
\hspace{0.5cm}
\includegraphics[scale=0.4]{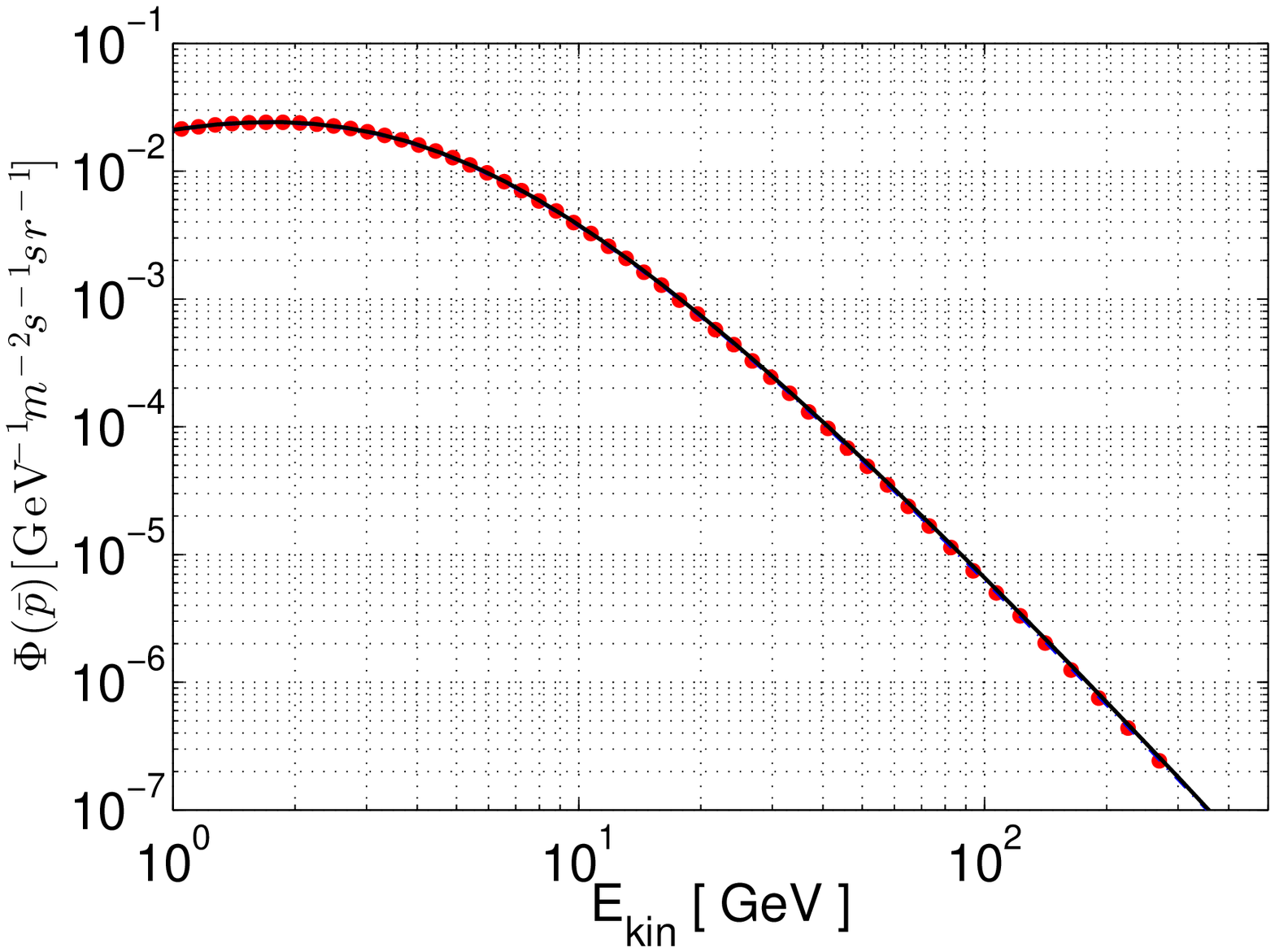}
\caption{\emph{{\small
Dark matter interpretation of the excess in \textsc{AMS-02}  positron fraction 
data \emph{(left panel)};
the corresponding flux of $\bar{p}$ would not show significant deviations
in \textsc{AMS-02}  $\bar{p}$ mock data \emph{(right panel)}, 
generated according what discussed in 
 Sect.~\ref{subsec:ams2}.
The dashed blue lines correspond to the contribution of the astrophysical backgrounds,
computed with method 2 described in Sect.~\ref{method2}.
 The solid black line is the total flux including the DM signal,
where we  have chosen  $\rm{DM \,DM }\to\tau^+\tau^-$,  $M_{\rm DM}= 1  \TeV$, $\sigma v=2.5\times 10^{-23} \,\rm{cm}^3 \rm{s}^{-1}$, Burkert halo profile.
 }}} 
\label{fig:data}
\end{figure}

The correlations we want to study in this paper are between positron fraction  and
$\bar{p}$ data, therefore the Cosmic-Ray (CR) species we have to deal with are $e^-, e^+, \bar p$. An example of how DM can account for the positron excess
without upsetting the $\bar{p}$ flux is shown in Fig.~\ref{fig:data}. 
In the following we will only consider
the data points with energies above 15 GeV, in order to be totally sure that the fluxes are not affected by solar modulation.
To generate the  signal fluxes at production, i.e. positrons and $\bar{p}$ from DM annihilations,
we use the numerical code PPPC4DMID described in Refs.~\cite{PPPC, dm1},
 which includes EW bremsstrahlung  from external legs in the factorization limit.
 The predictions of cosmic-ray fluxes originated from DM annihilations crucially depends
on how the DM is distributed in the galactic halo.
This astrophysical uncertainty is currently irreducible, therefore
we consider three of the usual profile choices:
\be
\rho(r)=\left\{
\begin{tabular}{cccl}
  $\displaystyle\rho_{s}\left[(1+r/r_{s})(1+ (r/r_{s})^{2})\right]^{-1}$, & 
 $r_s=12.67$ kpc, & $\rho_s=0.712$ GeV/cm$^3$,
 & (Burkert~\cite{Burkert})\\
 $ \rho_{s}\exp\left[-\frac{2}{0.17}\left[\left(r/r_{s}\right)^{0.17}-1\right]\right]$, & 
 $r_s=28.44$ kpc, & $\rho_s=0.033$ GeV/cm$^3$,
 & (Einasto~\cite{Einasto})\\
  $\displaystyle \rho_{s}(r_s/r)\left(1+r/r_{s}\right)^{-2}$, & 
 $r_s=24.42$ kpc, & $\rho_s=0.184$ GeV/cm$^3$,
 & (NFW~\cite{NFW})\\
\end{tabular}
\right.
\label{dmprofiles}
\ee
As for backgrounds, 
``primary'' electrons can come from galactic CR
while  interactions of CRs with the  interstellar medium is a source of ``secondary'' electrons,  positrons
and $\bar{p}$.
The propagation of the signal and background fluxes from their production region to the 
detector is affected by many processes, mainly diffusion and energy losses.
The number density per unit energy $n_i(r,z, p)$ of the cosmic ray species of interest $i$ $(= e^+,e^-, \bar p$),
 at any given point of cylindrical coordinates $r, z$  and  momentum $p$, evolves according to the transport equation
~\cite{Strong:1998pw}
\begin{eqnarray}
\frac{\partial n_i}{\partial t} &=& Q(r, z,p) + \nabla \cdot \left( D \nabla n_i  - \textbf{V}_c n_i \right) +
\frac{\partial }{ \partial p  } p^2 D_{pp} \frac{\partial }{ \partial p  } \frac{1}{p^2} n_i   \nonumber\\
&& - \frac{\partial }{ \partial p  } \left[\dot{p} n_i -\frac{p}{3} \left(\nabla \cdot \textbf{V}_c \right) n_i \right]  
- \frac{1}{\tau_{sp}}n_i -\frac{1}{\tau_f} n_i  \ ,
\label{eq:propeq}
\end{eqnarray}
where $D$ ($D_{pp}$) is the diffusion coefficient in position (momentum) space, 
$Q(r, z,p)$ is the source term due to DM annihilations, which depends on the DM mass, annihilation 
cross section and the energy spectrum at the interaction point (before propagation), 
$\textbf{V}_{c}$ is the convection velocity,  $\tau_{sp}$ is the 
time scale for nuclear spallation processes, 
$\tau_f$ is the time scale for fragmentation.
The flux  is then given by the solution of the transport equation through 
 $\Phi_i=\beta_i n_i/(4\pi)$, 
 where $\beta_i$ is the particle velocity in
units of the speed of light.

The coefficient $D_{pp}$ determines the diffusive reacceleration and is proportional to the the inverse of 
$D$ 
\begin{equation}  
D_{pp} = \frac{4}{3}\frac{p^2 v_A^2}{\delta ( 4- \delta^2) (4-\delta) } \frac{1}{D} \ ,
\end{equation}
where $v_A$ is the Alfv\'{e}n velocity. 
Eq.~(\ref{eq:propeq}) is solved in steady state ($\partial n/\partial t=0$), in a cylinder of height $2L$ and radius $R=20$ kpc, encompassing the galactic plane.
The diffusion coefficient is conventionally  parameterized  as $D(E)=\beta D_0 (\mathcal{R}/1 \rm{GV})^\delta$, where  
$\mathcal{R}$ its rigidity. Both the position dependence and the deviation from the linear dependence on $\beta$ have been neglected.
For the profile of the galactic magnetic field, which is rather uncertain,  we choose \cite{Strong:1998fr}
\be
B(r,z)=B_0 e^{-\frac{r-r_\odot}{10\, \textrm{kpc}}-\frac{|z|}{2\, \textrm{kpc}}}
\ee
where $r_\odot = 8.3$ kpc and $B_0$ is of the order of  $(1-10) \,\mu$G.

We carry out our analysis following two different methods for handling the propagation of signal and background
fluxes: in method 1, we take a conservative
approach letting  the astrophysical uncertainties largely vary; in method 2, we assume a specific propagation model 
and use it  to consistently propagate the signal and background fluxes of all relevant species.
Strictly speaking, the correct way to proceed would be to choose a specific CR propagation model within which to compute
both signal and background for all particle species and then scan over the model parameters.
We decided to adopt these two different methods because of simplicity and because
versions of each of them are often used in the literature.
They are  meant to represent two extreme ways of proceeding, so 
the ``truth'' is very likely to lie in between of the results obtained with
 these two methods.

\subsection{Method 1}
\label{subsec:method1}
 
The signal fluxes of $e^\pm$ and $\bar{p}$  due to DM annihilations, propagated to Earth, are computed with the
code PPPC4DMID, for different halo profiles and using 
a semi-analytic propagation model known as ``MED'' \cite{minmedmax, minmedmax2}.
This model corresponds to a specific choice of the coefficients of the diffusion-loss equation,  for
both $e^\pm$ and $\bar{p}$: $L=4$ kpc, $D_0=3.38 \times 10^{27} \textrm{cm}^2/\textrm{s}$, $\delta=0.70$, $V_{c}=12\,\rm{km/s}$, $V_A=52.9$ km/s, $B_0=4.78 \mu$G.

As a reference background $\bar{p}$ we use a spectrum  which 
is between the Kolmogorov (KOL) and Kraichnan (KRA) models of Ref.~\cite{Evoli:2011id} (the same choice has also been
done in Ref.~\cite{cirelliams}).
Changing the propagation model does not significantly affect the shape of the $\bar{p}$ flux,
but it changes the normalization \cite{Donato:2008jk}.
For primary electrons and secondary $e^\pm$, we use the parametrization \cite{Moskalenko:1997gh}, 
also used in \cite{Cirelli:2008id} (see instead Ref.~\cite{Meade:2009rb}
for a more data-driven approach)
\bea
\frac{\de \Phi_{e^-}}{\de E}&=&\left[ \frac{0.16 \epsilon^{-1.1}}{1+11\epsilon^{0.9}+3.2\epsilon^{2.15}}
+ \frac{0.7 \epsilon^{0.7}}{1+110\epsilon^{1.5}+600\epsilon^{2.9}+580 \epsilon^{4.2}}\right]
\,\textrm{GeV}^{-1}\,\textrm{cm}^{-2}\,\textrm{s}^{-1}\,\textrm{sr}^{-1}
\label{fluxem}\\
\frac{\de \Phi_{e^+}}{\de E}&=& \frac{4.5 \epsilon^{0.7}}{1+650\epsilon^{2.3}+1500\epsilon^{4.2}}
\,\textrm{GeV}^{-1}\,\textrm{cm}^{-2}\,\textrm{s}^{-1}\,\textrm{sr}^{-1}
\label{fluxep}
\eea
where $\epsilon=E/$ GeV.
Each of these background functions suffer from astrophysical uncertainties
related to the CR propagation in our Galaxy.
We deal with theses uncertainties 
by simply allowing the backgrounds to have floating normalizations
and slopes; in practice, we
multiply each of the background fluxes by  $A_i \, \epsilon^{p_i}$ ($i=e^-, e^+, \bar p$), 
and marginalizing over $A_i\in[0.5,2]$ and  $p_i\in[-0.05,0.05]$.

This approach has the virtue of accounting for the astrophysical uncertainties of the  propagation model in a simple and conservative way.
It will therefore result in rather cautious bounds on the parameter space.
On the other hand, 
 the relationship between the $A,p$ and the physical parameters of the transport equation is obscured. Furthermore, 
 the marginalization over $A,p$ parameters is
 done independently for $e^\pm$ and $\bar{p}$. This corresponds to treat the  normalizations 
 and slopes of the backgrounds of  these types of fluxes as uncorrelated, as they are
 propagated with different physical parameters.

\subsection{Method 2}
\label{method2}

\begin{figure}[t]
\centering
\includegraphics[scale=0.4]{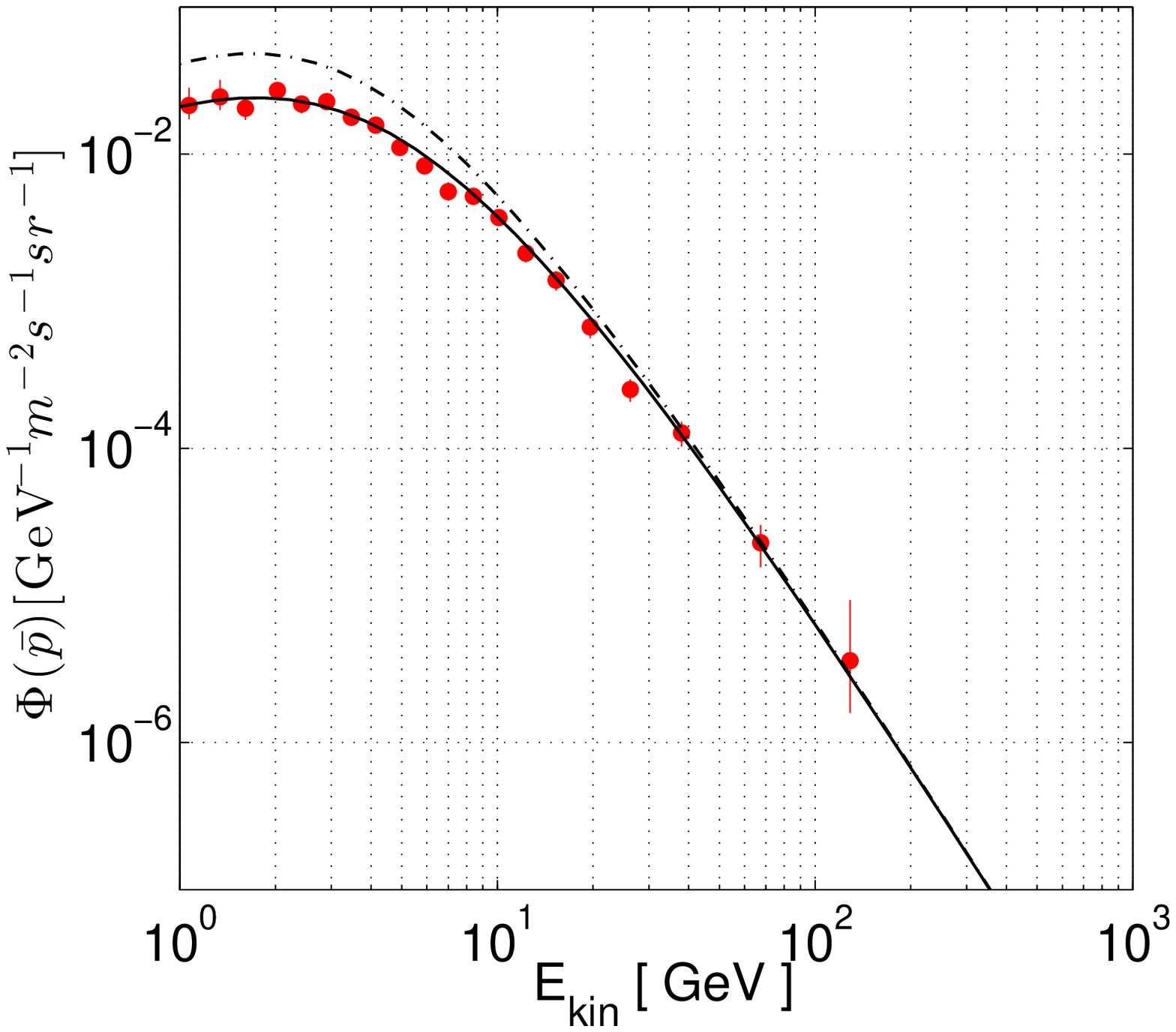}
\hspace{1cm}
\includegraphics[scale=0.4]{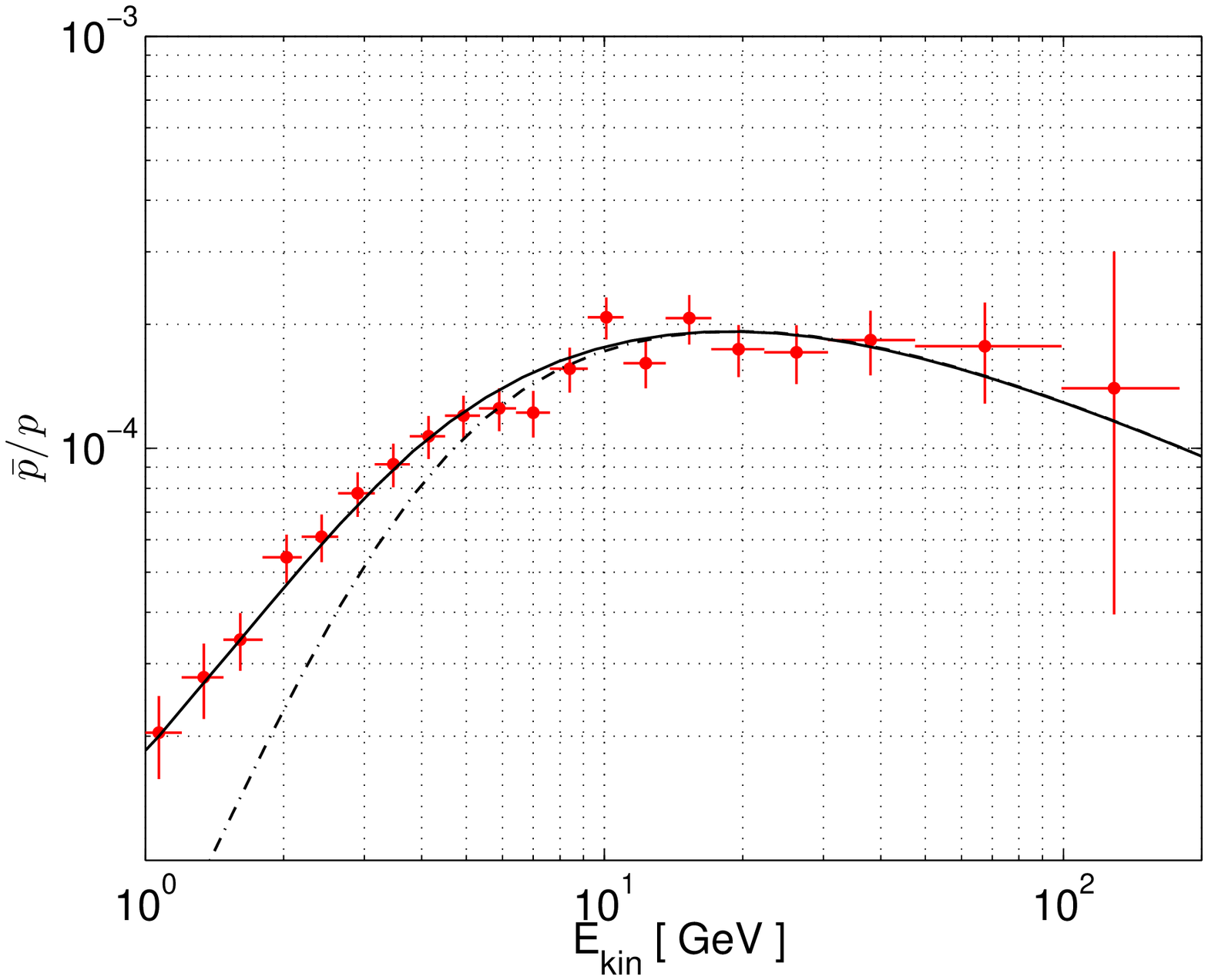}
\caption{\emph{{\small
Fits of our reference propagation model, as defined in Sect.~\ref{method2}, to current data: 
\textsc{Pamela} $\bar{p}$  \emph{(left panel)}, 
\textsc{Pamela} $\bar{p}/p$ \emph{(right panel)}.
Dot-dashed lines refers to the case without correcting for solar modulation effects.
}}}
\label{fig:ourmodelfits}
\end{figure}

Alternatively, we consider another way to deal with astrophysics.
We insist on the fact that we want to  consider a consistent
setup where both signal and background are propagated with the same set of propagation 
parameters.
So we consider a specific propagation model  defined by 
$L=4$ kpc, $D_0=1.50 \times 10^{28} \textrm{cm}^2/\textrm{s}$, $\delta=0.50$, $V_A=13.715$ km/s, $B_0=7.0 \mu$G.
Unlike the MIN, MED, MAX models, no convection  is considered.
We set the  injection index of primary protons to be 2.35, while the injection index  of primary electron spectra is 0.6 (2.53) below (above) the reference rigidity of 4 GV. 
This model provides a good fit to\cite{FermiLAT:2011ab}, \textsc{Pamela} positron 
\cite{Adriani:2008zr} and $\bar{p}$  \cite{Adriani:2010rc} data
(see Fig.~\ref{fig:ourmodelfits}), as well as to  \textsc{Fermi} all-electron data at low energy  and B/C data.

The signal fluxes due to DM annihilations are computed
by computing the energy spectra at the production point with the PPPC4DMID
code, and then propagating them using GALPROP \cite{galprop} and the model specified above.
The background fluxes for $e^{\pm}$ and $\bar{p}$ are also computed using GALPROP, 
 for the same propagation model used for the signal. 

Notice that, differently from method 1, for method 2 we assume we have a rather precise knowledge
of the astrophysical details. As of today, this is not the case, but for instance
the future data releases of \textsc{AMS-02}  on absolute $e^+, e^-, p, \bar p$ as well as  light nuclei ratios, 
will improve the accuracy of the  propagation parameters and of the resulting backgrounds
(see e.g. Refs.~\cite{minmedmax2, Delahaye:2008ua}).
Therefore, carrying on this method serves as a representative way for what can be achieved
with more detailed information about CR propagation in our Galaxy.

This approach has  the virtue of propagating consistently all fluxes within the same model,
although it pays the price of not being generic, as a scan over the uncertainties of the
model parameters is not performed.

\section{Results}
\label{sec:results}

\subsection{Analysis of the positron fraction data}

We first perform a $\chi^2$ analysis of the \textsc{AMS-02} positron fraction  data
for the different channels under consideration
$e^+e^-, \mu^+ \mu^-, \tau^+\tau^-$,
in order to identify the best-fit one.
We also checked  that best-fit regions obtained by fitting the positron signal with the 
$b\bar b$ and $W^+ W^-$  channels are already exlcuded by \textsc{Pamela} anti-$p$ data \cite{Adriani:2010rc}.
As already mentioned, we restrict ourselves to use only the $E \gtrsim 15$ GeV bins of the data, as the lower energy bins are strongly affected by solar modulation effects, so a total of 36 data points.
The numbers of degrees of freedom 
are 36-6=30 and  36-2=34, for method 1 and 2, respectively.

\begin{figure}[t!]
\centering
\begin{tabular}{cc}
\textbf{Method 1}
&\hspace{1cm}
\textbf{Method 2}\\
\includegraphics[scale=0.4]{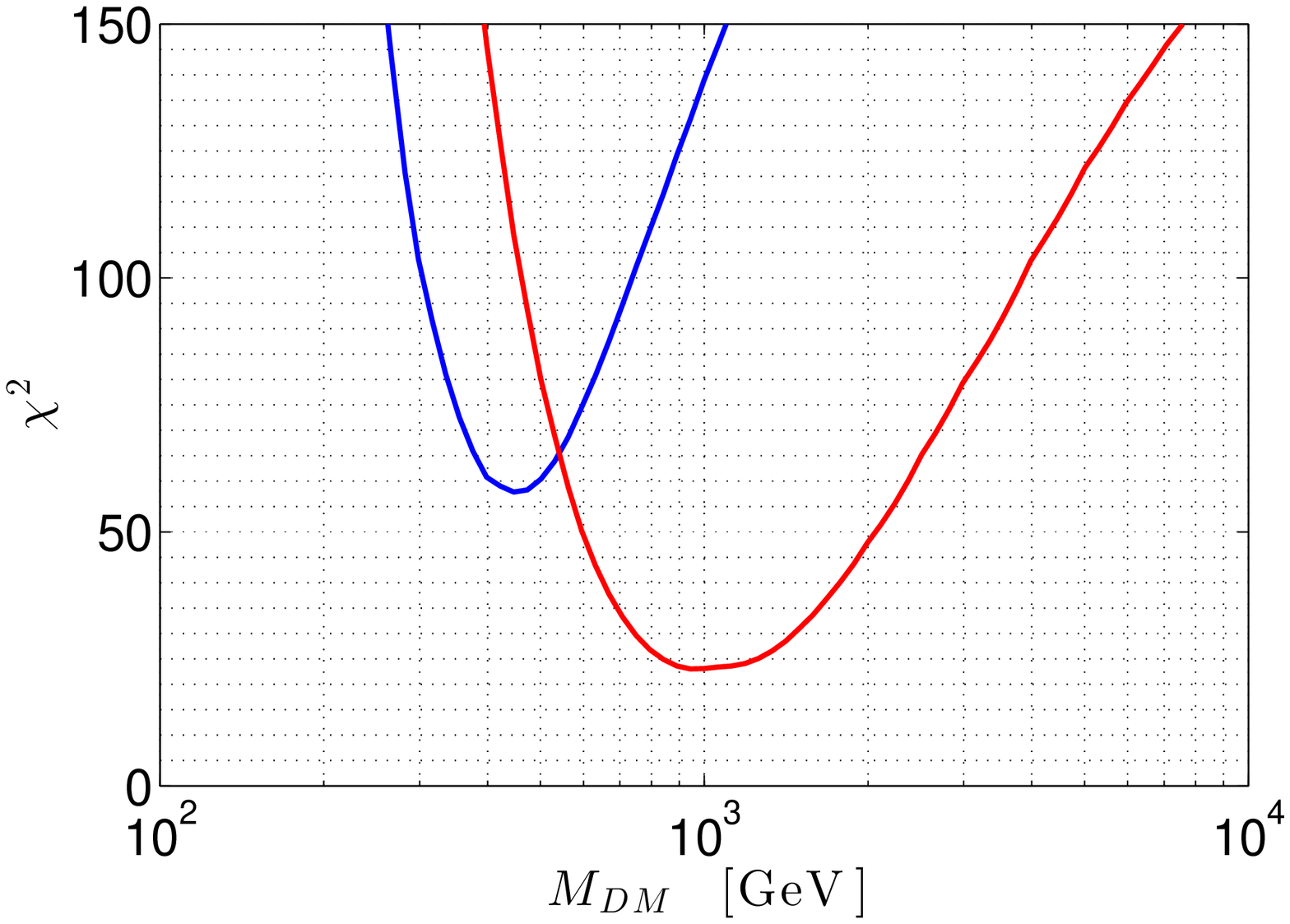}
&\hspace{0.5cm}
\includegraphics[scale=0.4]{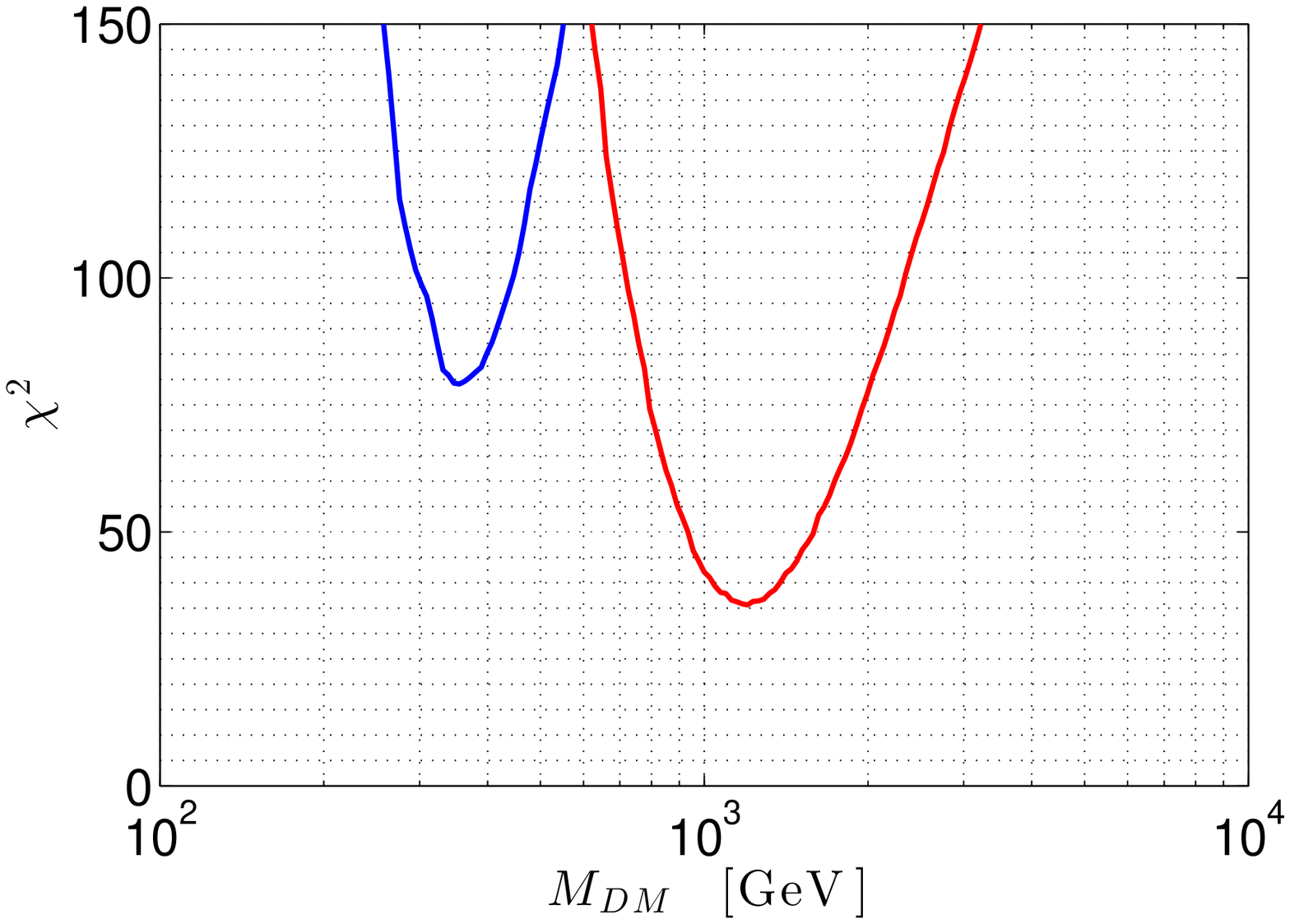}\\
\end{tabular}
\caption{\emph{{\small
The $\chi^2$ function as a function of the DM mass, 
of the \textsc{AMS-02}
positron fraction data, 
for method 1 \emph{(left panel)}  and 2 \emph{(right panel)}.
The halo profile is Einasto.
Different colours correspond to different DM annihilation channels:
$\mu^+\mu^-$ \emph{(blue)}, $\tau^+\tau^-$ \emph{(red)}.
The $e^+e^-$ channel
gives an even higher $\chi^2$, not shown in the plot.
}}}
\label{fig:chisquare}
\end{figure}

In Fig.~\ref{fig:chisquare} we show the $\chi^2$  as a function of the mass, 
minimized over the annihilation cross section, 
for the $\mu$ and $\tau$ channels.
As a reference halo profile we chose Einasto, but for  other choices 
the results are very similar.
The $e$ channel is not shown as it gives a very poor fit with even higher values of 
$\chi^2$: $\chi^2/\rm{dof}\gtrsim 6.6, 5.8$, for methods 1 and 2, respectively.
The $\mu$ channel gives  $\chi_{\rm min}^2/\rm{dof}\simeq 1.9, 2.4$ for methods 1 and 2, respectively, so it  also provides a rather poor fit to the data.
Instead, the $\tau$ channel gives  $\chi_{\rm min}^2/\rm{dof}\simeq 0.7, 1.0$ for methods 1 and 2,
respectively.
So we conclude that the only case producing a good fit to the 
\textsc{AMS-02} positron fraction 
data are  DM of about 1 TeV annihilating into  $\tau^+\tau^-$.
We also note 
that the case of DM annihilation into  $\mu^+\mu^-$, although it gives higher $\chi^2$ to
 \textsc{AMS-02} positron data than $\tau^+\tau^-$,
 it is less  constrained by gamma-ray measurements  \cite{Ackermann:2012rg}.

After having identified the situation which fits the signal at best, 
we now want to proceed and correlate it with future measurements
of anti-$p$ flux by the same experiment. To this end, we  
 simulate the detector response and generate mock data.

\begin{figure}[t!]
\centering
\begin{tabular}{cc}
\textbf{Method 1}
&\hspace{1cm}
\textbf{Method 2}\vspace{0.5cm}\\
\includegraphics[scale=0.4]{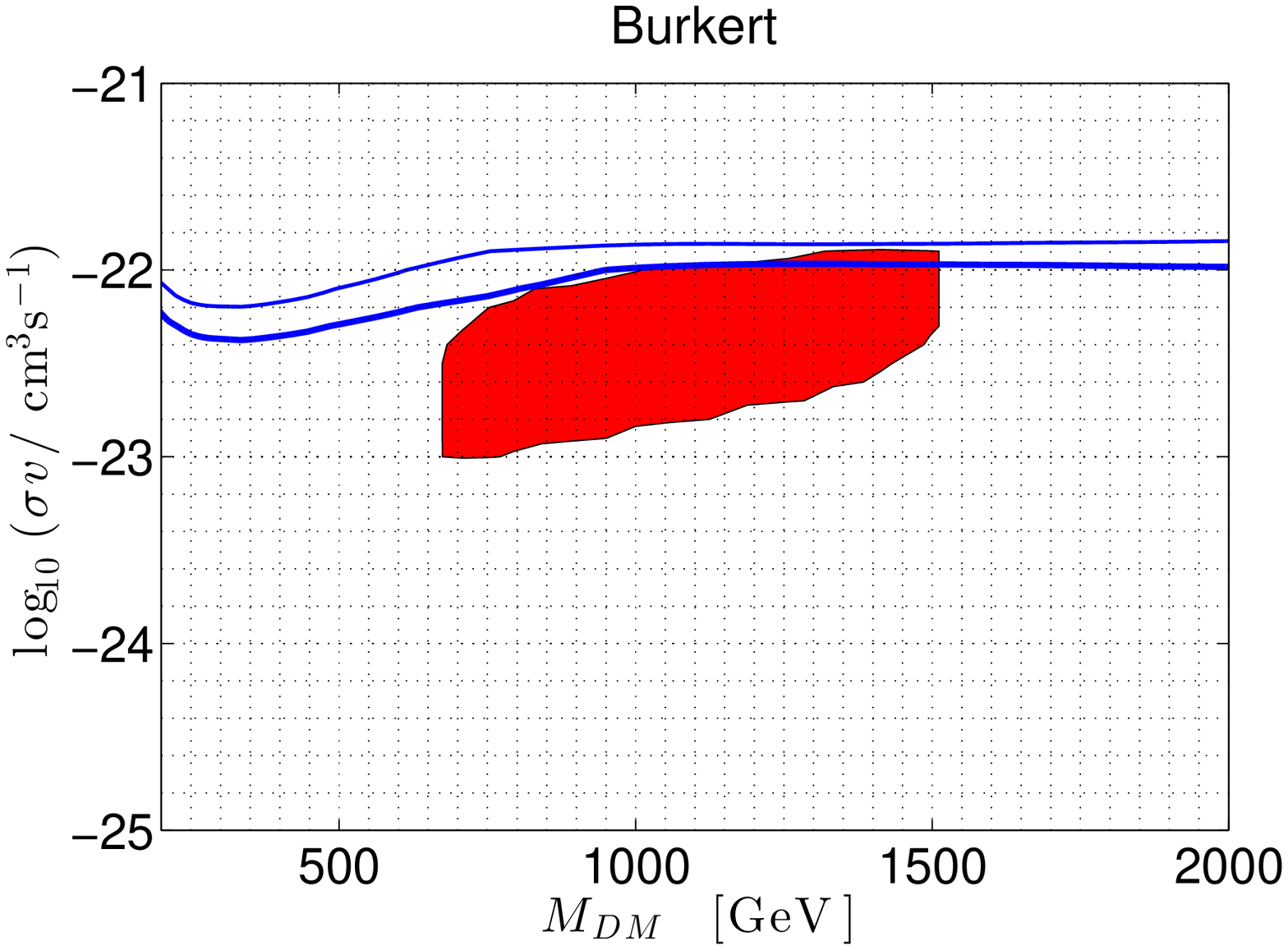}
&\hspace{1cm}
\includegraphics[scale=0.4]{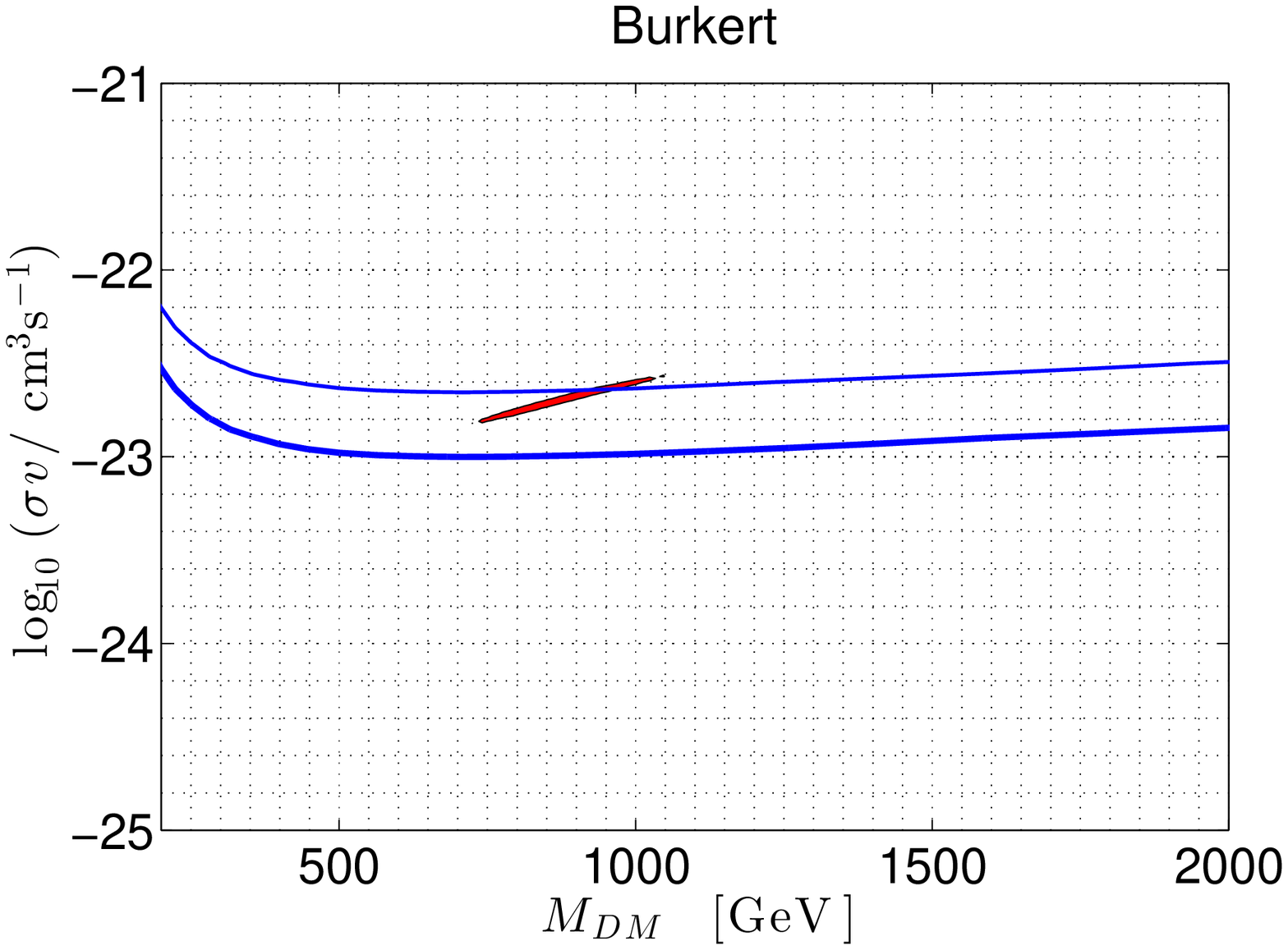}\\
\includegraphics[scale=0.4]{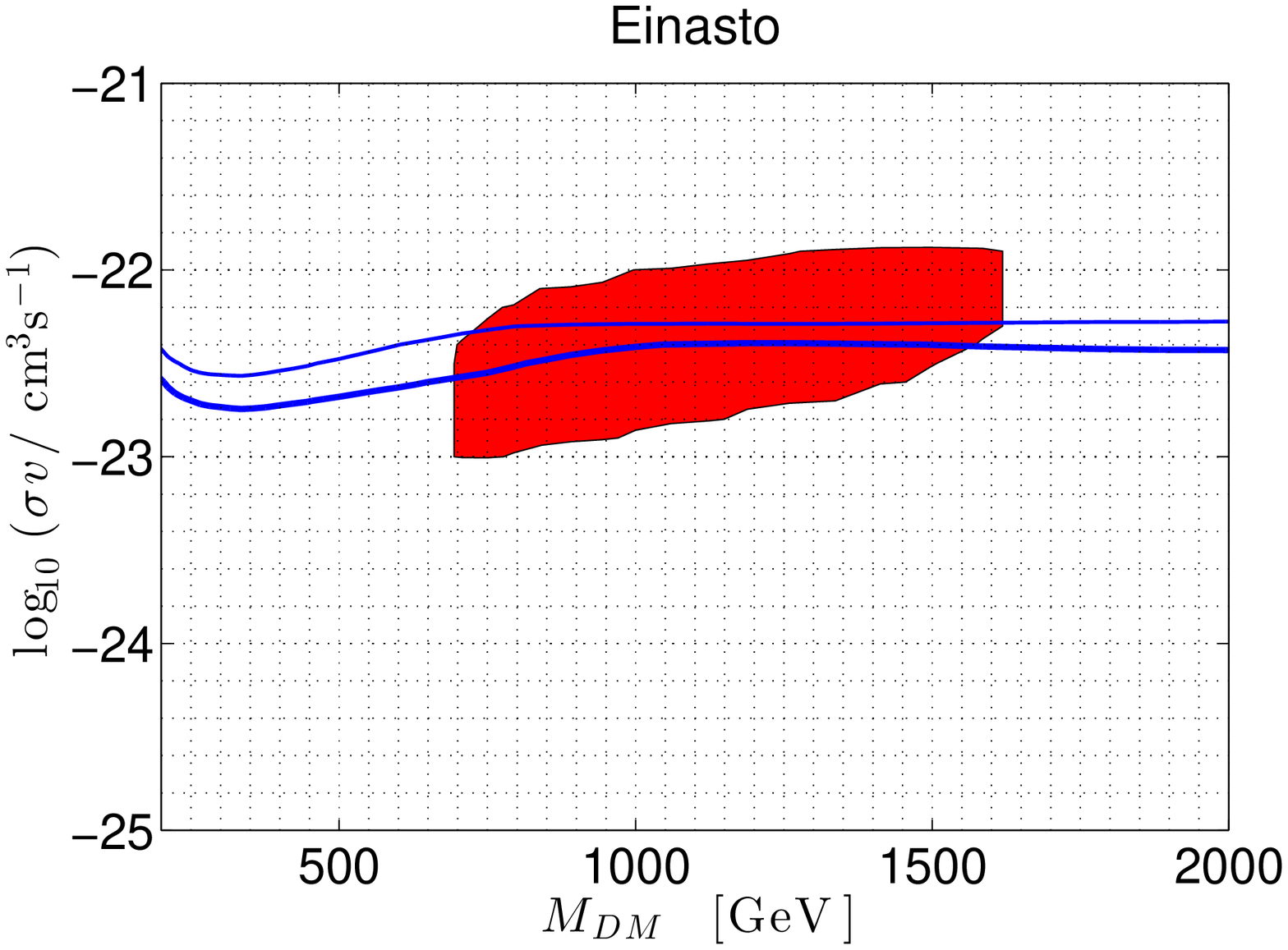}
&\hspace{1cm}
\includegraphics[scale=0.4]{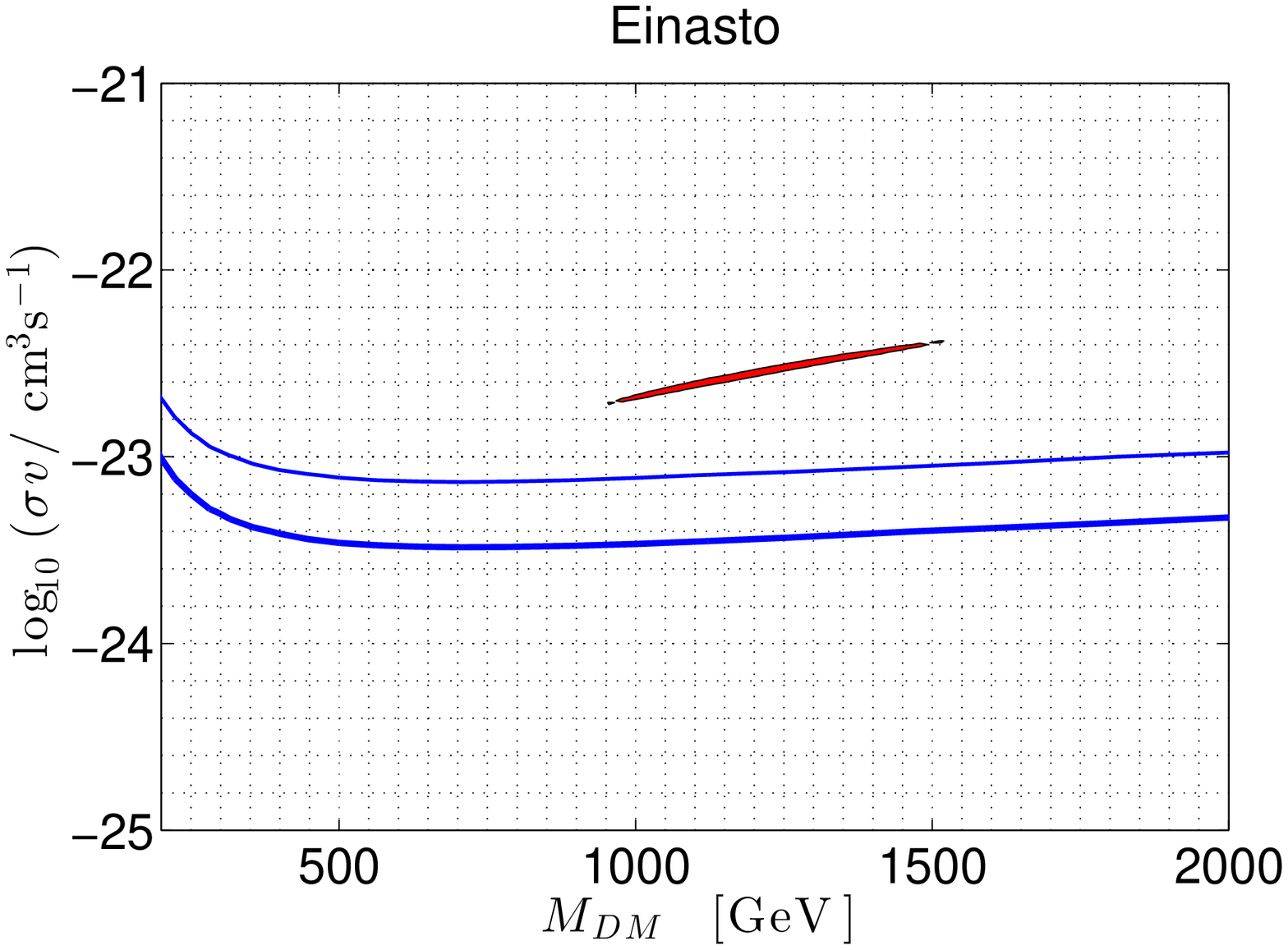}\\
\includegraphics[scale=0.4]{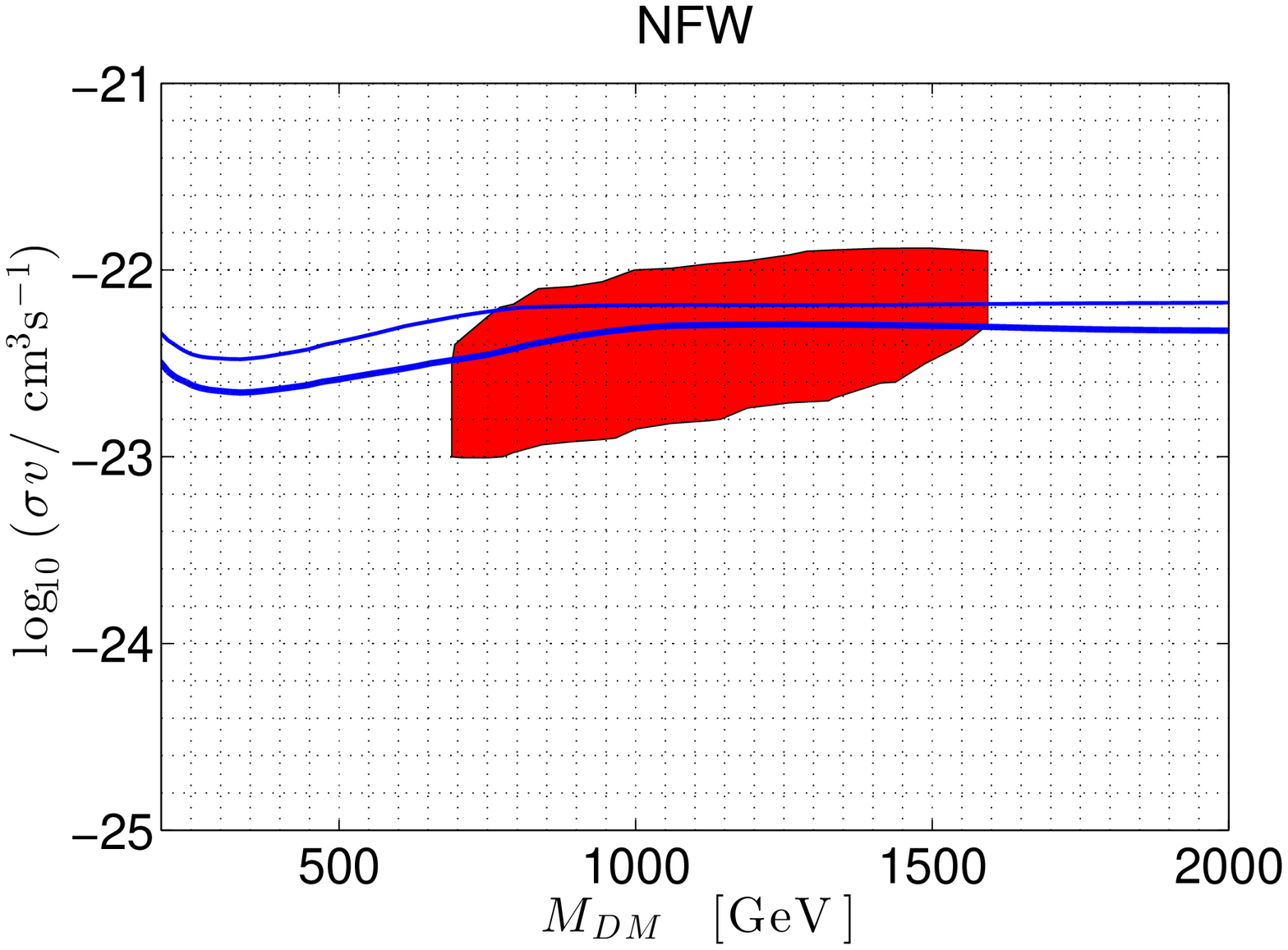}
&\hspace{1cm}
\includegraphics[scale=0.4]{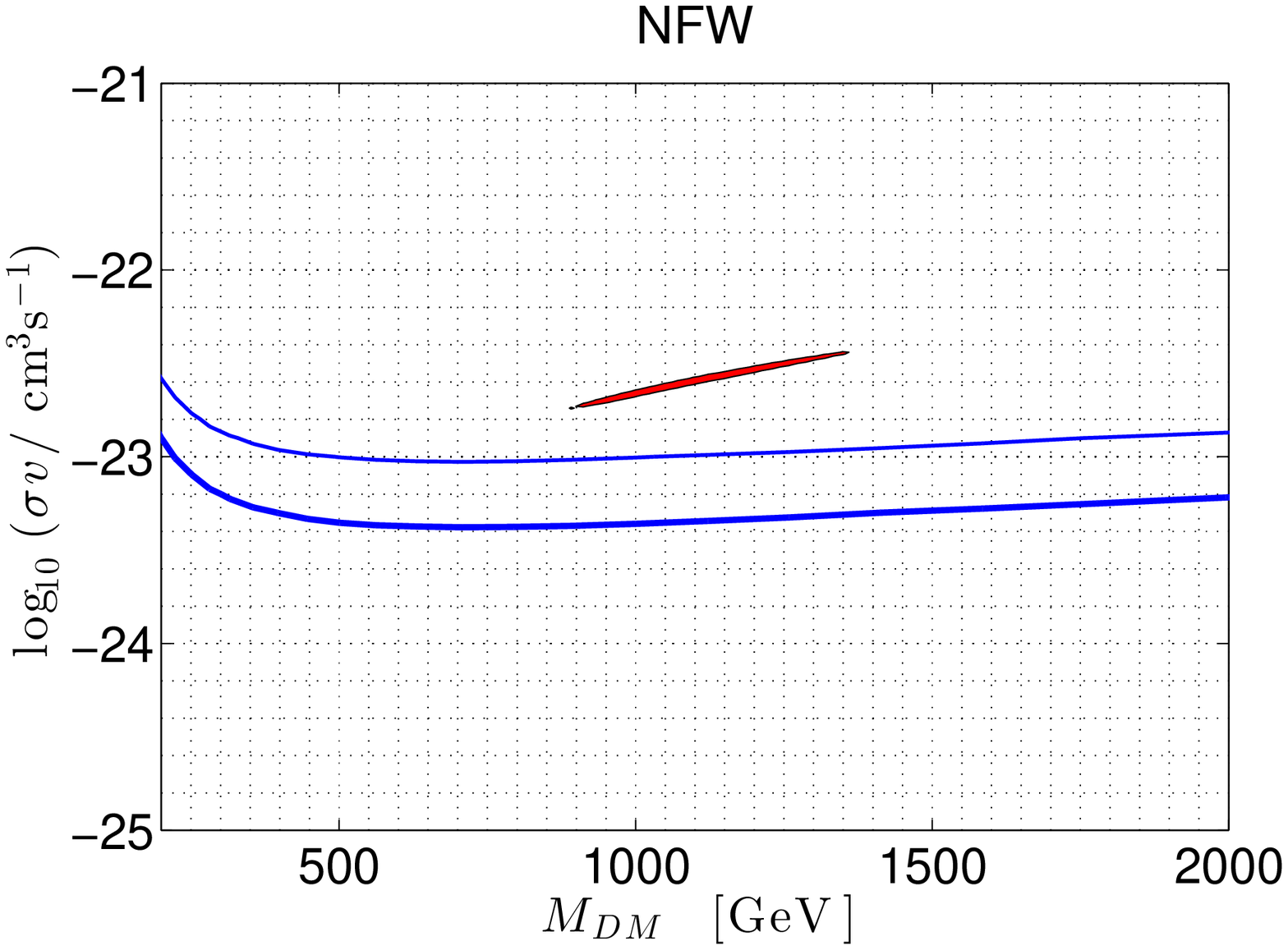}
\end{tabular}
\caption{\emph{{\small
The red contours are the best-fit $3\sigma$ regions of the parameter space $(m_\chi, \langle\sigma v\rangle)$,
of the \textsc{AMS-02}
positron fraction data with {\rm DM DM} $\to \tau^+\tau^-$ channel, 
for method 1 \emph{(left column)}  and method 2 \emph{(right column)}, 
as described in the text.
The  blue  lines are the projected exclusions (at 99.7\% CL) due to $\bar{p}$ \textsc{AMS-02} mock data,
for 1-year \emph{(thin blue line)} and 3-year of data taking \emph{(thick blue line)}.
From top to bottom: Burkert,  Einasto, NFW profiles. 
}}}
\label{fig:fitstau}
\end{figure}

\subsection{Simulation of \textsc{AMS-02} $\bar{p}$ data}
\label{subsec:ams2}

As we have mentioned in the Introduction, we work under the hypothesis that the future  data on the $\bar{p}$ flux
reported by \textsc{AMS-02} will not show any evidence of DM annihilations and will be therefore compatible with the 
expected background. In order to assess the projected sensitivity of   \textsc{AMS-02}, we assume that the $\bar{p}$ flux will be measured up to energies of the order of 300 GeV and that the number of detected $\bar{p}$ in a given kinetic energy bin  $(E_{\rm kin}-\Delta E_{\rm kin}/2, E_{\rm kin}+\Delta E_{\rm kin}/2)$ is
\be
N_{\bar p}(E_{\rm kin})\simeq  a_{\bar p}(E_{\rm kin})\Phi_{\bar p}(E_{\rm kin})\Delta E_{\rm kin} \Delta t\,,
\ee
being   $a_{\bar p}$ the geometrical acceptance, $\Phi_{\bar p}$ the $\bar{p}$ flux and $\Delta t$ the exposure time, and we have taken 100\% efficiency \cite{efficiency}. 
Assuming a Poisson distribution, the statistical error is therefore $(\Delta N_{\bar p})_{\rm stat}/{N_{\bar p}}=
1/\sqrt{N_{\bar p}}$.
For the systematic error for $\bar{p}$ detection and reconstruction,
we can only suppose the
syst error is in the 1-10\% range \cite{hooper};
therefore we consider as reasonable values $(\Delta N_{\bar p})_{\rm syst}/{N_{\bar p}}=5\%$ for the 1-year data and $(\Delta N_{\bar p})_{\rm syst}/{N_{\bar p}}=3\%$ for the 3-year data,
to be added in quadrature to the statistical error.
For the geometric acceptance of $\bar{p}$ we choose  $a_{\bar p}=0.2$ m$^2$ sr as a safe lower
limit above 1 GeV  
 \cite{pohl}.
Finally,  we use a linear approximation of the detector energy 
resolution (see \cite{tingtalk, cirelliams}), $\Delta E_{\rm kin}/E_{\rm kin}=(0.042\cdot (E_{\rm kin}/\GeV)+10)\%$, which allows about 20 bins per decade.

Within this setup, we generated mock $\bar{p}$ data for \textsc{AMS-02} in the range $(10-300)$ GeV of kinetic
energies, following the two reference backgrounds discussed in the previous section. 
In the right panel of Fig.~\ref{fig:data} we show the mock data generated as described above, as well as the reference
background and the DM contribution to the $\bar{p}$ flux, for method 2.

\subsection{Correlations between $e^+$ and $\bar p$ data}

For the best-fit situation DM DM $\to \tau^+\tau^-$ we  find 
the $3\sigma$ best-fit contours, on the $(m_{\rm DM}, \langle\sigma v\rangle)$ plane, corresponding to $\Delta \chi^2<11.8$, of the \textsc{AMS-02}
positron fraction data.
They are shown as red contours in  Fig.~\ref{fig:fitstau},
 for the two methods discussed 
in the previous section.
On the same plots, the  blue  lines are the projected exclusions (at 99.7\% CL) due to $\bar{p}$ \textsc{AMS-02} mock data,
for 1-year and 3-year of data taking, generated according what discussed in 
the subsection \ref{subsec:ams2}, and assuming
the data will not show deviations from the background.
We checked that varying the geometrical acceptance $a_{\bar p}$ by a factor of 2 does not change the results significantly, as the errors are dominated by systematics.
We vary the DM profile, following the choices in Eq.~(\ref{dmprofiles}), and we use the two different methods described in Section \ref{sec:background}.
We reiterate that significant portions of the parameter space shown in Fig.~\ref{fig:fitstau} are already 
excluded by complementary data, such as $\gamma$-rays \cite{Ackermann:2012rg}.
However, the bounds for shallow halo profiles like isothermal or Burkert are 
weaker.
We do not include these constraints in our plots as the focus of this paper is on
the correlations between \textsc{AMS-02} current and projected data.

The results are not altered significantly by choosing  Einasto or NFW halo profiles, 
but exclusions are weaker for the Burkert profile, as it assumes lower DM density near the  center, and therefore a lower  $\bar{p}$ signal would be produced.

Furthermore, the results are sensitive to the method chosen for handling the propagation of signal and background.
The fact that the best-fit contours of the positron fraction data are  smaller for method 2 than for method 1
is not a surprise, as in method 2 we have  less freedom to vary the astrophysical background
and therefore the DM signal is more constrained.
For the same reason, the exclusion curves coming from $\bar{p}$ projected data are more stringent
for method 2.
Therefore,  adopting method 1, any conclusion is weaker than for method 2, and we can imagine the actual situation will lie in between of these two scenarios.

We can conclude  that 1-year of \textsc{AMS-02} data on $\bar{p}$ would already be enough to almost rule out the DM interpretation of the positron excess.
With 3 years of data all the exclusion curves are obviously more stringent.

The high energy bins of the \textsc{AMS-02} positron fraction data show a decrease in the slope and suggest a possible turning down of the  signal.
If  future positron data releases will  show more clearly a fall of the positron fraction, 
the best-fit contours with DM signal would shrink and better pin down the DM mass.

\section{Conclusions}
\label{sec:conclusions}

This paper  demonstrates the relevance of correlations induced by the EW corrections in indirect DM searches.
The results of the analysis described above allow us to draw the following  conclusions:

\begin{itemize}
\item if the recently released \textsc{AMS-02} positron fraction data are interpreted
as a signal of annihilating DM, then the $\chi^2$ analysis favours a 
a DM of about 1 TeV annihilating into $\tau^+\tau^-$
(a situation which is already severely constrained by $\gamma$-ray measurements);

\item assuming no signal is seen in future $\bar{p}$ data released by \textsc{AMS-02}, one will be able to exclude almost completely the DM interpretation of the positron excess, by  using only data from a single experiment. 

\end{itemize}
These conclusions have to be considered together with
two \textit{caveat}s. First,
while the energy spectra of SM particles  at the interaction point  coming from DM are predictable in terms of particle physics, the propagation of these particles
in the complex galactic environment  is still uncertain. 
Therefore,  astrophysical uncertainties and the way to deal with them affect
the results.
We have pointed out the importance to use consistently the same propagation setup for both signal and background,
which is especially relevant when combining together the fluxes of more than one particle species.
Second, although the assumptions we made are consistent with the understanding of the  \textsc{AMS-02} detector from outside
the collaboration, 
our results may be sensitive to the detector features we used for generating the projected data.

This work should be regarded as  the first of a program where we will investigate more aspects and examples of 
correlations among DM signals. 
In subsequent publications we will study different channels and update these results as soon as new
data become available.

\section*{Acknowledgments}
We thank M.~Pohl for many helpful discussions on the \textsc{AMS-02} detector features and
P.~Serpico for useful conversations on the lepton signals from pulsars.
We also thank D.~Gaggero and A.~Strumia for interesting conversations.



{\small
\begin{thebibliography}{99}


\bibitem{review} For a recent review of the state-of-the-art, see e.g. M.~Cirelli,
  Pramana {\bf 79}, 1021 (2012)
    	  \href{http://arXiv.org/abs/1202.1454}{[arXiv:1202.1454]}.

\bibitem{ams02positron} S. Ting, ``Recent results from the  \textsc{AMS} experiment", seminar  given at CERN on 3$^{\rm rd}$ April 2013;
M. Aguilar et al. (AMS Collaboration), Phys. Rev. Lett. \textbf{110}, 141102 (2013) 
  
\bibitem{dm1}
  P.~Ciafaloni, D.~Comelli, A.~Riotto, F.~Sala, A.~Strumia and A.~Urbano,
  JCAP {\bf 1103}, 019 (2011)
	  \href{http://arXiv.org/abs/1009.0224}{[arXiv:1009.0224]}.
    
\bibitem{dm2}
  P.~Ciafaloni, M.~Cirelli, D.~Comelli, A.~De Simone, A.~Riotto and A.~Urbano,
  JCAP {\bf 1106}, 018 (2011) 
  \href{http://arXiv.org/abs/1104.2996}{[arXiv:1104.2996]}.

\bibitem{dm3}
  P.~Ciafaloni, M.~Cirelli, D.~Comelli, A.~De Simone, A.~Riotto, A.~Urbano,
  JCAP {\bf 1110}, 034 (2011)
   \href{http://arXiv.org/abs/1107.4453}{[arXiv:1107.4453]}.
 
\bibitem{dm4}  
  P.~Ciafaloni, D.~Comelli, A.~De Simone, A.~Riotto and A.~Urbano,
  JCAP {\bf 1206}, 016 (2012)
   \href{http://arXiv.org/abs/1202.0692}{[arXiv:1202.0692]}.  
 
       
\bibitem{dm5} 
N.~F.~Bell, J.~B.~Dent, T.~D.~Jacques and T.~J.~Weiler,
  Phys.\ Rev.\  D {\bf 83}, 013001 (2011)
                 \href{http://arXiv.org/abs/1009.2584}{[arXiv:1009.2584]}.

\bibitem{dm6}                  
 N.~F.~Bell, J.~B.~Dent, T.~D.~Jacques and T.~J.~Weiler,
  Phys.\ Rev.\ D {\bf 84}, 103517 (2011)
                   \href{http://arXiv.org/abs/1101.3357}{[arXiv:1101.3357]}.

\bibitem{dm7} 
  N.~F.~Bell, J.~B.~Dent, A.~J.~Galea, T.~D.~Jacques, L.~M.~Krauss and T.~J.~Weiler,
  Phys.\ Lett.\ B {\bf 706}, 6 (2011)
  \href{http://arXiv.org/abs/1104.3823}{[arXiv:1104.3823]}.

\bibitem{Adriani:2008zr} 
  O.~Adriani {\it et al.}  [PAMELA Collaboration],
  Nature {\bf 458}, 607 (2009)
	  \href{http://arXiv.org/abs/0810.4995}{[arXiv:0810.4995]};
  O.~Adriani, G.~C.~Barbarino, G.~A.~Bazilevskaya, R.~Bellotti, M.~Boezio, E.~A.~Bogomolov, L.~Bonechi and M.~Bongi {\it et al.},
  Astropart.\ Phys.\  {\bf 34}, 1 (2010)
	  \href{http://arXiv.org/abs/1001.3522}{[arXiv:1001.3522]}.

\bibitem{FermiLAT:2011ab} 
  M.~Ackermann {\it et al.}  [Fermi LAT Collaboration],
  Phys.\ Rev.\ Lett.\  {\bf 108}, 011103 (2012)
	  \href{http://arXiv.org/abs/1109.0521}{[arXiv:1109.0521]}.    

\bibitem{Adriani:2010rc} 
  O.~Adriani {\it et al.}  [PAMELA Collaboration],
  Phys.\ Rev.\ Lett.\  {\bf 105}, 121101 (2010)
  	  \href{http://arXiv.org/abs/1007.0821}{[arXiv:1007.0821]}.    

\bibitem{bertone} M.~Pato, M.~Lattanzi and G.~Bertone,
  JCAP {\bf 1012}, 020 (2010)
  	  \href{http://arXiv.org/abs/1010.5236}{[arXiv:1010.5236]}.    

\bibitem{inprep} A. De Simone {\it et al.}, in preparation.

\bibitem{Meade:2009iu} 
  P.~Meade, M.~Papucci, A.~Strumia and T.~Volansky,
  Nucl.\ Phys.\ B {\bf 831}, 178 (2010)
   \href{http://arXiv.org/abs/0905.0480}{[arXiv:0905.0480]}.    
 
 
\bibitem{Cirelli:2009dv} 
  M.~Cirelli, P.~Panci and P.~D.~Serpico,
  Nucl.\ Phys.\ B {\bf 840}, 284 (2010)
   \href{http://arXiv.org/abs/0912.0663}{[arXiv:0912.0663]}.      
  
\bibitem{PPPC} 
  M.~Cirelli, G.~Corcella, A.~Hektor, G.~Hutsi, M.~Kadastik, P.~Panci, M.~Raidal and F.~Sala {\it et al.},
  JCAP {\bf 1103}, 051 (2011)
  	  \href{http://arXiv.org/abs/1012.4515}{[arXiv:1012.4515]}.

\bibitem{Burkert}
A.~Burkert,
  IAU Symp.\  {171} (1996) 175
  [Astrophys.\ J.\  {447} (1995) L25]
                   \href{http://arXiv.org/abs/astro-ph/9504041}{[astro-ph/9504041]};
  
\bibitem{Einasto}
  A.~W.~Graham, D.~Merritt, B.~Moore, J.~Diemand and B.~Terzic,
  Astron.\ J.\  {132} (2006) 2685
                         \href{http://arXiv.org/abs/astro-ph/0509417}{[astro-ph/0509417]};
  J.~F.~Navarro {\it et al.},
                           \href{http://arXiv.org/abs/0810.1522}{arXiv:0810.1522}.

\bibitem{NFW}
  J.~F.~Navarro, C.~S.~Frenk and S.~D.~M.~White,
  Astrophys.\ J.\  {462} (1996) 563
                       \href{http://arXiv.org/abs/astro-ph/9508025}{[astro-ph/9508025]}.
                                      
\bibitem{Strong:1998pw} 
  V.~L.~Ginzburg, (ed.), V.~A.~Dogiel, V.~S.~Berezinsky, S.~V.~Bulanov and V.~S.~Ptuskin,
  Amsterdam, Netherlands: North-Holland (1990) 534 p;
  A.~W.~Strong and I.~V.~Moskalenko,
  Astrophys.\ J.\  {\bf 509}, 212 (1998)
    	  \href{http://arXiv.org/abs/astro-ph/9811296}{[astro-ph/9807150]}.    

\bibitem{Strong:1998fr} 
  A.~W.~Strong, I.~V.~Moskalenko and O.~Reimer,
  Astrophys.\ J.\  {\bf 537}, 763 (2000)
  [Erratum-ibid.\  {\bf 541}, 1109 (2000)]
    	  \href{http://arXiv.org/abs/astro-ph/9811296}{[astro-ph/9811296]}.  
  


  \bibitem{minmedmax}
  F.~Donato, N.~Fornengo, D.~Maurin and P.~Salati,
  Phys.\ Rev.\ D {\bf 69}, 063501 (2004)
  	  \href{http://arXiv.org/abs/astro-ph/0306207}{[astro-ph/0306207]}.
 
  \bibitem{minmedmax2}  
  T.~Delahaye, R.~Lineros, F.~Donato, N.~Fornengo and P.~Salati,
  Phys.\ Rev.\ D {\bf 77}, 063527 (2008)
  	  \href{http://arXiv.org/abs/0712.2312}{[arXiv:0712.2312]}.    
  
  
\bibitem{Evoli:2011id} 
  C.~Evoli, I.~Cholis, D.~Grasso, L.~Maccione and P.~Ullio,
  Phys.\ Rev.\ D {\bf 85}, 123511 (2012)
  	  \href{http://arXiv.org/abs/1108.0664}{[arXiv:1108.0664]}.  
  
  \bibitem{cirelliams} 
  M.~Cirelli and G.~Giesen,
  	  \href{http://arXiv.org/abs/1301.7079}{arXiv:1301.7079}.



\bibitem{Donato:2008jk} 
  F.~Donato, D.~Maurin, P.~Brun, T.~Delahaye and P.~Salati,
  Phys.\ Rev.\ Lett.\  {\bf 102}, 071301 (2009)
  \href{http://arXiv.org/abs/0810.5292}{[arXiv:0810.5292]}.
  


\bibitem{Moskalenko:1997gh} 
  I.~V.~Moskalenko and A.~W.~Strong,
  Astrophys.\ J.\  {\bf 493}, 694 (1998)
  	  \href{http://arXiv.org/abs/astro-ph/9710124}{[astro-ph/9710124]};  
  E.~A.~Baltz and J.~Edsjo,
  Phys.\ Rev.\ D {\bf 59}, 023511 (1998)
  	  \href{http://arXiv.org/abs/astro-ph/9808243}{[astro-ph/9808243]}.


\bibitem{Cirelli:2008id} 
  M.~Cirelli, R.~Franceschini and A.~Strumia,
  Nucl.\ Phys.\ B {\bf 800}, 204 (2008)
   \href{http://arXiv.org/abs/0802.3378}{[arXiv:0802.3378]}.        


\bibitem{Meade:2009rb} 
  P.~Meade, M.~Papucci and T.~Volansky,
  JHEP {\bf 0912}, 052 (2009)
  	  \href{http://arXiv.org/abs/0901.2925}{[arXiv:0901.2925]}.  

 \bibitem{galprop} 
	  \href{http://galprop.stanford.edu/}{http://galprop.stanford.edu/}.  
   

\bibitem{Delahaye:2008ua} 
  T.~Delahaye, F.~Donato, N.~Fornengo, J.~Lavalle, R.~Lineros, P.~Salati and R.~Taillet,
  Astron.\ Astrophys.\  {\bf 501}, 821 (2009)
   \href{http://arXiv.org/abs/0809.5268}{[arXiv:0809.5268]}.          
 
\bibitem{Ackermann:2012rg} 
M.~Ackermann {\it et al.}  [LAT Collaboration],
  Astrophys.\ J.\  {\bf 761}, 91 (2012)
  \href{http://arXiv.org/abs/1205.6474}{[arXiv:1205.6474]}.
  
\bibitem{efficiency}
A. Oliva, 
High Charge Cosmic Rays Measurement with
the AMS-02 Silicon Tracker,
PhD thesis, 2007.

\bibitem{hooper}   M.~Pato, D.~Hooper and M.~Simet,
  JCAP {\bf 1006}, 022 (2010)  	  
  \href{http://arXiv.org/abs/1002.3341}{[arXiv:1002.3341]}.

\bibitem{pohl}
M.~Pohl, private communication.

\bibitem{tingtalk}
S.~Ting, slides of the talk at 
\href{http://indico.cern.ch/internalPage.py?pageId=1&confId=197799}{SpacePart12},
5-7 November 2012, CERN.    


\end{thebibliography}
}
\end{document}